\documentstyle[12pt,epsfig]{article}
\newcommand{\be}{\begin{equation}}
\newcommand{\ee}{\end{equation}}

 1
\font\elevenrm=cmr10 scaled\magstep 1
 1

\def\reference{\hang\noindent}

\def\gtorder{\mathrel{\raise.3ex\hbox{$>$}\mkern-14mu
             \lower0.6ex\hbox{$\sim$}}}
\def\ltorder{\mathrel{\raise.3ex\hbox{$<$}\mkern-14mu
             \lower0.6ex\hbox{$\sim$}}}

\newcommand{\tacc}{t_{\rm acc}}
\newcommand{\tcross}{t_{\rm cr}}
\newcommand{\tc}{t_{\rm c,3}}
\newcommand{\tvar}{t_{\rm var}}
\newcommand{\tflare}{t_{\rm f}}
\newcommand{\fflare}{\eta_{\rm f}}

\newcommand{\tesc}{t_{\rm esc}}

\newcommand{\elcomp}{\ell_{\rm e}}

\newcommand{\gammamax}{\gamma_{\rm max}}

\newcommand{\diff}{{\rm d}}

\newcommand{\eqb}{\begin{eqnarray}}
\newcommand{\eqe}{\end{eqnarray}}

\begin{document}

\vspace*{1.8cm}
  \centerline{MODELS OF VARIABILITY IN BLAZAR JETS}
\vspace{1cm}
  \centerline{A. MASTICHIADIS$^1$, J.G. KIRK$^2$}
\vspace{1.4cm}
  \centerline{$^1$PHYSICS DEPARTMENT, UNIVERSITY OF ATHENS}
  \centerline{\elevenrm GR-15784 Zografos, Athens, Greece}
\vspace{1.4cm}
  \centerline{$^2$MAX-PLANCK-INSTITUT F\"UR KERNPHYSIK}
  \centerline{\elevenrm Postfach 10 39 80, D-69029 Heidelberg, Germany}
\vspace{3cm}
\begin{abstract}
We present the expected variability features in the context of the 
Synchro-Self-Compton (SSC) model of emission from 
Active Galactic Nuclei (AGNs).
We show that the homogeneous SSC model can describe well the
observed multiwavelength spectrum and variability of objects such as Mkn 421
and Mkn 501; however in order to explain the very fast
flaring timescales occasionally observed from 
these objects one should use a laminar, rather than spherical,
source geometry. This picture leads naturally to a shock-in-jet model
which we approach as a site of diffusive particle acceleration.
We show that this acceleration scheme can explain certain characteristic 
features of AGN X-ray spectra and can provide us with further observational
tests.
\end{abstract}

\vspace{2.0cm}

\section {Introduction}

During the last decade there has been a drastic change in our picture 
of high energy emission from Active Galactic Nuclei (AGNs). 
The 2nd CGRO catalogue lists over 40 AGNs as strong sources of 
GeV gamma rays (Thompson et al 1995) while two AGNs, Mkn 421 and 
Mkn 501, have been detected in the TeV regime (Punch et al. 1992,
Quinn et al. 1996). All of these 
AGNs belong to the category of blazars, which include OVV, flat radio 
sources, many of which exhibit superluminal motion. The fact that up 
to date not a single radio quiet AGN has been detected in the GeV/TeV
regime (Lin et al. 1992) 
has put a first strong constraint on the theories of gamma-ray emission 
from AGNs, while, at the same time, has given arguments in favour of certain
theories for AGN unification (Urry \& Padovani 1995). 

On the theoretical front it became quickly apparent that the gamma-ray 
emission was connected with processes in the jet rather than in the core.
While this general picture remains more or less undisputed, many
models have been proposed for the high energy emission itself; these
can be roughly divided in leptonic or hadronic in origin, depending on
whether it is electrons or protons which are responsible for the gamma-ray
emission.  
Thus while there are models which invoke protons as the ultimate source 
of high energy emission (Mannheim 1993 , Bednarek \& Protheroe 1997a), the majority 
of the proposed models assume that the gamma-rays come from inverse 
Compton scattering of relativistic electrons on some soft photon targets.
The source of these targets is still an open question and many possible origins 
have been proposed such as accretion disk photons 
(Dermer, Schlickeiser \& Mastichiadis 1992), diffuse isotropic photons 
coming from regions such as the broad line clouds (Sikora, 
Begelman \& Rees 1994), internally produced synchrotron photons 
(Maraschi, Ghisellini \& Celotti et al. 1992; Marscher \& Travis 1996,
Inoue \& Takahara 1996) or combinations thereof (Dermer, Sturner, 
Schlickeiser 1997)
with each model giving rather similar spectral features 
and characteristics.

A very interesting aspect which emerged from the intense 
gamma-ray monitoring of the sources was the discovery of fast 
variability. So in addition to the already known variability in the X-ray regime, 
Mkn 421 was discovered to exhibit TeV flares, the fastest of which
had a duration of about 
15 minutes (Gaidos et al. 1996). More powerful sources, such as 3C279, 
have shown variability in the GeV regime of the order of an hour (Hartman 
et al. 1996). These observations put new, interesting constraints on the 
theoretical models of high energy emission from AGNs since one 
expects the particle cooling times to be of the order of the flare itself. 
The imposed constraints become even tighter from recent results of 
multiwavelength campaigns which show certain trends in the evolution of 
flares along the EM spectrum. Thus Mkn 421 was discovered to exhibit 
quasisimultaneous variation in the keV and TeV regime (Macomb et al. 1995), 
while other energy regimes (most notably the GeV regime) remained 
virtually unaffected.
The other AGN detected in TeV, Mkn~501, has shown similar trends (Catanese
et al. 1997, Pian et al. 1998).

The aforementioned observations provoked a flurry of models which addressed 
explicitly either the fast variability (Salvati, Spada \& Pacini 1998), the multiwavelength 
spectrum (Ghisellini, Maraschi \& Dondi 1997) or both (Mastichiadis \& Kirk 1997). In 
\S 2 of
the present article we will review the basic features of such models 
especially in the context of the so-called homogeneous synchrotron 
self-Compton models (SSC).
In \S 3 we will address explicitly the problem of particle acceleration 
and we will present a simple way one can explain certain 
observations with the picture of accelerating/radiating particles. 

\section{Homogeneous Synchrotron-Self Compton Models}

\subsection{Spherical Models}
This class of models, 
based on the ideas first put forward by Jones, O'Dell \& Stein (1974),
has extensively been discussed elsewhere (Kirk
\& Mastichiadis 1997,
Mastichiadis \& Kirk 1997--henceforth MK97, and Ghisellini, Maraschi \& Dondi 1997),
however for the sake of completeness we give a brief overview here. 
As the above authors
have shown, a homogeneous region containing magnetic fields and 
relativistic electrons can reproduce the observed spectrum of the blazar 
Mkn 421 whilst allowing for time variations on the scale of 
roughly 1 day. In order to address explicitly the temporal behaviour of the spectrum,
MK97 used a set of time-dependent, 
spatially averaged kinetic equations for the electrons and photons 
adopting the approach outlined in Mastichiadis \& Kirk (1995).  
The electrons are assumed to have a power-law uniform injection in a spherical
source  (blob) of radius $R$; the blob itself is supposed to move at some small 
angle $\theta$ to our line of sight with a bulk Lorentz factor $\Gamma$. 
The electrons lose energy from synchrotron radiation on a 
magnetic field of strength $B$ and from inverse Compton 
radiation on the produced synchrotron photons.
The so obtained electron distribution function is then convolved with 
the single electron synchrotron and inverse Compton emissivities and 
the overall photon spectrum is obtained after allowing for the 
possibility of photon-photon pair production --a process which turns out to be 
negligible for the parameters used. 

Seven independent parameters are needed to determine a stationary 
spectrum in this model. They are the Doppler-boosting factor 
$\delta=[\Gamma(1-B_{\rm b}\cos\theta)]^{-1}$ (with $B_{\rm b}c$ the bulk velocity 
of the source), the size of the source $R$, its magnetic field $B$, 
the mean time during which particles are confined in the source $\tesc$, 
and three parameters determining the injected relativistic electron 
distribution: its luminosity, or compactness $\elcomp$, the spectral index 
$s$ and the maximum Lorentz factor of the electron distribution $\gammamax$.
The inclusion of the particle escape time $\tesc$ becomes 
necessary from the fact that the photon spectrum is rather flat between 
the radio and the infra-red region, implying that the radiating particles 
do not have time to cool significantly. It turns out that this fit leaves a  
free parameter which can be suitably chosen as either the Doppler 
factor of the blob $\delta$ or the timescale over which variability 
can be observed $\tvar$ (in sec). These two quantities are related by the scaling 
relation $\delta=267\tvar^{-1/4}$ ($\tvar$ expressed in sec). 
For the reported variability of about 
one day ($\tvar=10^5$ sec--Macomb et al 1995) one readily finds 
$\delta=15$, which is close to the usually assumed values of the Doppler 
factor. The spectrum of the flare can be fitted
in a time-dependent fashion (i.e. before complete cooling can be 
achieved) by changing $\gammamax$ by a factor of a few (MK97). 

\subsection{Slab models} 

Recently, very rapid variations in the TeV flux of Mkn~421 have been 
reported (Gaidos et al 1996). As it was stated above, in the framework 
of the homogeneous SSC spherical models this implies that acceptable 
fits can be found only by increasing the Doppler factor $\delta$. 
Thus, as the above scaling formula between $\delta$ and $\tvar$
suggests, a choice of 
$\tvar=1000$ sec (so as to agree with the observed flare timescale) 
would mean $\delta$ of the order of 50, a value which is above 
those indicated by observations of apparent superluminal motion 
(Vermeulen \& Cohen 1994). 

\begin{figure}[t]
\vspace{- .5 cm}
\centerline{\epsfxsize=11. cm
\epsffile{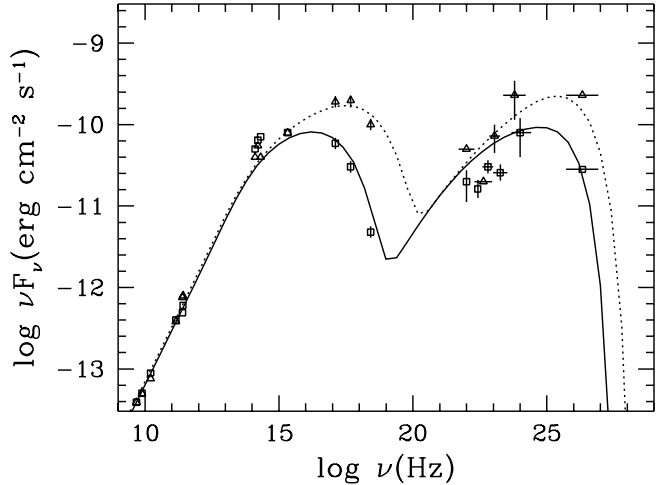}}
\vspace{-1.5 cm}
\caption{Low and high states of the multiwavelength spectrum of 
Mkn421. The data points were taken from Macomb et al (1996). The fits were
made using a laminar geometry for the emitting region. For the parameters
used see text.}  
\end{figure}

An alternative way of approach can be understood as follows:
Assuming that the electrons cool due to synchrotron radiation
and that their cooling time $\tc$ is given in units of $10^3$ sec
we can write $\tc\simeq 10^6\gamma^{-1}B^{-2}\delta^{-1}$ sec where
$\gamma$ is the Lorentz factor of the particle in the rest frame of
the emitting plasma and $B$ is the magnetic field in gauss.
The highest energy photons (in units of 10 keV) emitted by these particles 
are $\nu_{10}\simeq 10^{-12}\gamma^2B\delta$.
These relations imply $B\simeq 1\tc^{-2/3}\nu_{10}^{-1/3}\delta^{-1/3}$
gauss and $\gamma\simeq 10^6\tc^{1/3}\nu_{10}^{2/3}\delta^{-1/3}$.
Therefore the maximum photon energy radiated by such electrons
is $\nu_{\rm max}\simeq .5\tc^{1/3}\nu_{10}^{2/3}\delta^{2/3}$ TeV.
Assuming furthermore that the source has equal amounts of energies
in magnetic fields and photons (as seems to be implied by the source's 
equal amounts of luminosities in the radio to X-ray  and soft to very
high energy gamma-rays regimes), we obtain an expression for the aspect ratio of
the source region $\eta=d/R$ where $d=3~10^{13}\delta\tc$ cm is a thickness
measured in the rest frame of the source and $R$ is defined such that $\pi R^2$
is the area of the source when projected onto the plane of the sky.

\begin{figure}[t]
\centerline{\epsfxsize=11. cm
\epsffile{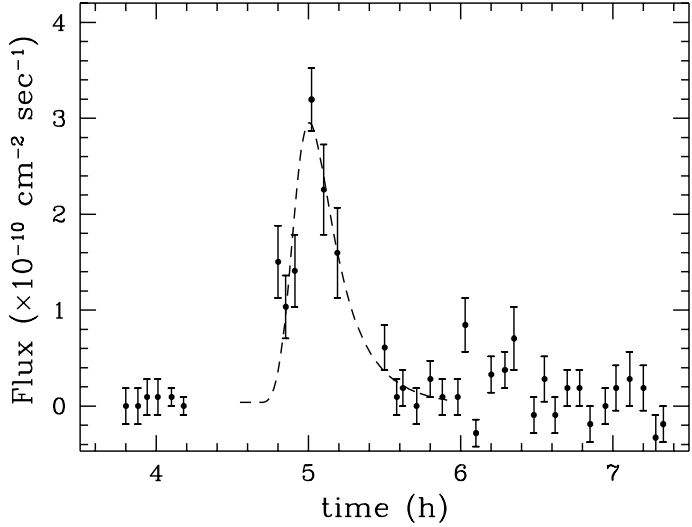}}
\vspace{-2.5 cm}
\caption{Fit to the flare of Mkn421 observed by Gaidos et al. (1996) for the 
the parameters used in producing the multiwavelength fit of 
Figure 1. The flare was obtained by  increasing the amplitude of the
injected electrons within one crossing time by a factor of 12.}
\end{figure}

Fig.1 shows a fit to the multiwavelength spectrum of Mkn 421 as this
was given in Macomb (1995, 1996). A fit to the same data
can be obtained as well by assuming a spherical source but
with either a long variability timescale $\tvar$ and a `canonical' 
Doppler factor $\delta$ or with a short $\tvar$ and a high $\delta$ 
(see MK97). The present fit for the low state
was obtained for $\tvar=500~s$,
$\delta=20$, $B=0.4$ G, $\gammamax=1.4~10^5$, $s=1.7$, $\elcomp=1.5~10^{-5}$
and $\tesc=50\tcross$. The high state was obtained by increasing
$\gammamax$ by a factor of 4 while leaving the other 
parameters unchanged.
The corresponding values of $d$ and $R$ are $9.2~10^{13}$ cm and $2.8~10^{15}$ cm
respectively, implying an aspect ratio of $\eta\simeq .03$.

As it was pointed in MK97 (and can also be seen from Fig.~1) changes 
in $\gammamax$ result in large variations in the X and
TeV regime but these changes are not especially prominent at lower frequencies. 
An alternative way of producing a flare is to consider an increase
in the amplitude $Q$ of the injected relativistic electrons
while leaving the other parameters unchanged.
Figure 2 shows the TeV flare produced by increasing $Q$ by a factor
of 12 within one crossing time and consequently decreasing it to its
original value. The produced flare can fit quite well the flare
reported by Gaidos et al. (1996). 

Figure 3 shows the corresponding low and high states of Mkn~501.
The parameters used are 
$\tvar=500~s$,
$\delta=15$, $B=0.5$ G, $\gammamax=1.6~10^5$, $s=1.8$, $\elcomp=.7~10^{-5}$
and $\tesc=100\tcross$. The high state was obtained by increasing
$\gammamax$ by a factor of 40 (to accommodate the fact that Mkn 501
during one flare in 1997
was observed by OSSE up to energies of 200 keV-- Pian et al. 1998).
The corresponding values of $d$ and $R$ are $7.1~10^{13}$ cm and $1.1~10^{16}$ cm
respectively, implying an aspect ratio of $\eta\simeq .007$.

\begin{figure}[t]
\hspace{2 cm}
\epsfxsize=9.5 cm
\epsffile{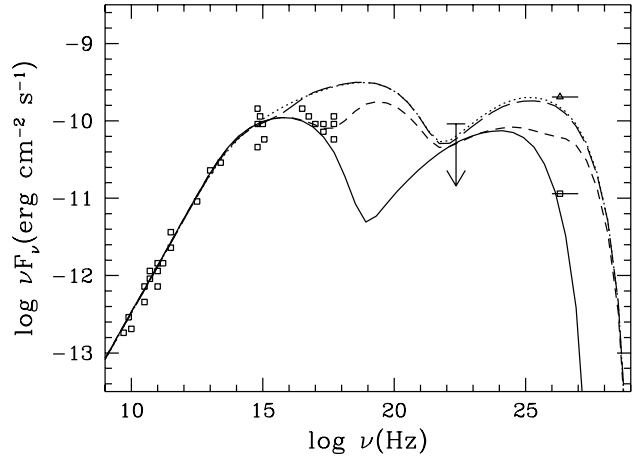}
\vspace{-1.5 cm}
\caption{\protect\label{multi501}
Low and high states of the multiwavelength spectrum of Mkn 501 for a 
laminar emitting region. Data points were taken
from Catanese et al. (1997) and Pian et al.(1998). The short and long 
dashed lines correspond to the spectrum 3 and 30 $t_{\rm cr}$ respectively
after the change in $\gammamax$. The dotted line corresponds to the new steady-state.}
\end{figure}

From the above it is evident that fast variability can be accommodated
in the homogeneous self-Compton models only in the case where the 
emitting source is a thin structure with a crossing time comparable
to the cooling time of the highest energy particles. This can lead 
naturally to the shock-in-jet model (Marscher \& Gear 1985) which
we present in the next section. 

\section{Particle Acceleration in Blazar Jets}

Let us consider a thin shock
wave moving down a cylindrically symmetric jet (Marscher \& Gear 1985, 
Kirk, Rieger, \& Mastichiadis 
1998--henceforth KRM) with a velocity $u_{\rm s}$ in the rest frame of the jet. 
Let also
particles be accelerated by the shock through a first order Fermi scheme and 
subsequently escape downstream where they radiate. Following KRM we will
restrict the present analysis only to synchrotron losses and radiation. 

The equation that governs the number of particles $N(\gamma)$
with Lorentz factors between $\gamma$ and $\gamma+d\gamma$ in the acceleration zone
can be written
  \eqb
\label{phanomen}
   \frac{\partial N}{\partial t} + \frac{\partial}{\partial \gamma} 
\left[\left(
   {\gamma\over\tacc}  -\beta_{\rm s}\,\gamma^2 \right) N \right] +
   {N\over \tesc}  &=& Q \delta (\gamma - \gamma_0)
   \eqe
(Kirk, Melrose \& Priest 1994),
   where 
   \eqb
   \beta_{\rm s}&=& 
\frac{4}{3}\frac{\sigma_{\rm T}}
    {m_{\rm e} c^2}  
    \left(\frac{B^2}{8\pi}\right) \,.
   \eqe                 
with $\sigma_{\rm T}$ the
Thomson cross section. The first term in brackets in 
the above equation describes acceleration at the
rate $\tacc^{-1}$, 
the second describes the rate of energy loss due to synchrotron radiation 
averaged over pitch-angle 
(because of the assumed isotropy of the distribution) 
in a magnetic field $B$. Particles 
are assumed to escape from this region at an energy independent rate 
$\tesc^{-1}$, and to be injected into the acceleration 
process with a (low) Lorentz factor $\gamma_0$ at a rate $Q$ particles per 
second. Note that the concept of this \lq acceleration zone\rq\ differs
from the emission region in the homogeneous model discussed in the previous
section in two important
respects: a) particles are injected at low energy and continuously accelerated and
b) very little radiation is emitted by a particle whilst in the acceleration
zone. A further difference comes from the fact that the high energy cut-off of the
electron distribution is given now by a detailed balance between the acceleration
and loss rates at the Lorentz factor $\gamma_{\rm max}=1/(\beta_{\rm s}\tacc)$.
The variability features 
therefore do not depend only on the electron cooling timescale but on
the interplay between  acceleration and loss timescales.
For $\gamma<\gammamax$ the acceleration rate exceeds the synchrotron loss rate while for
$\gamma>\gammamax$ the distribution vanishes.

To describe the kinetic equation in the radiation zone we follow Ball \& Kirk (1992) and
use a coordinate system at rest in the radiating plasma. 
The shock front then 
provides a moving source of electrons, which subsequently suffer energy 
losses, but are assumed not to be transported in space. The kinetic 
equation governing the differential 
density $\diff n(x,\gamma,t)$ of particles in 
the range $\diff x$, $\diff \gamma$ is then
\eqb
\label{trans}
\frac{\partial n}{\partial t} -
\frac{\partial}{\partial \gamma}(\beta_{\rm s}\,\gamma^2\,n) &=&
{N(\gamma,t)\over\tesc}\delta(x-x_{\rm s}(t)) 
\eqe
where $x_{\rm s}(t)$ is the position of the shock front at time $t$.
For a shock which starts to accelerate (and therefore \lq 
inject\rq) particles at time $t=0$ and position $x=0$ and moves
at constant speed $u_{\rm s}$, the solution of Eq.~(\ref{trans}) 
for $\gamma>\gamma_0$ is 
\eqb
n(x,\gamma,t)&=&
{a\over u_{\rm s}\tesc\gamma^2}
\nonumber\\
&&\left[\frac{1}{\gamma}-\beta_{\rm s} \left(t -
  \frac{x}{u_{\rm s}}\right) - 
\frac{1}{\gammamax}\right]^{(\tacc-\tesc)/\tesc}
\nonumber\\
&&
\Theta \left[\gamma_1(x/u_{\rm s})- 
   (1/\gamma - \beta_{\rm s}t+ \beta_{\rm s}x/u_{\rm s})^{-1}\,\right]
\,,
\label{fulldist}
\eqe
where $\gamma_1(t)$ is given by 
\eqb
\gamma_1(t) &=& \left(\frac{1}{\gammamax} + 
\left[\frac{1}{\gamma_0} - 
   \frac{1}{\gammamax}\right]\,{e}^{-t/\tacc}\right)^{-1}\,.
\label{gamma1eq}
\eqe

\begin{figure}[t]
\vspace{-1.5 cm}
\centerline{\epsfxsize=9. cm
\epsffile{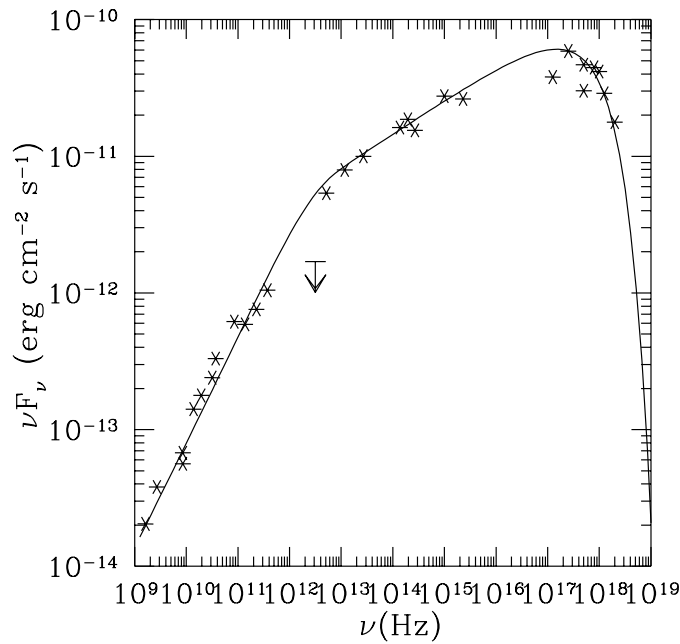}}
\vspace{-.5 cm}
\caption{The radio--X-ray spectrum of Mkn 501 as calculated from the acceleration
model described in Section 3 once in steady-state. The data were taken from the
collation of Catanese et al. 1997.} 
\end{figure}

To obtain the synchrotron emissivity as a function of position, time and
frequency we convolve the density $n$ with the
  synchrotron Green's function $P(\nu,\gamma)$. 
At a point  $x=X$ ($>u_{\rm s}t$)
on the symmetry axis 
of the source at time $t$
the specific intensity of radiation in the $\vec{x}$ direction 
depends on the retarded time ${\bar t}=t-X/c$ and is given by
\eqb
\label{jetframe}
I(\nu,{\bar t})\,=\,\int\diff\gamma P(\nu,\gamma)\int\diff x
\,n(x,\gamma,{\bar t}+x/c)
\eqe
At this point we stress that in this model one needs to integrate
the differential electron density over the spatial coordinate
since, in contrast to the homogeneous models, the acceleration
region is distinct from the cooling region.

\subsection{Spectral Signatures of Acceleration}

As in the case of the homogeneous models we first seek parameters
that could fit specific blazar spectra in a steady state and
then we try to induce a flare by changing some parameter of the
fit. 

As an example, we show in 
Fig.~4 observations of the object Mkn~501.
The gamma-ray emission of this object is not included in this
figure, since, according to \S 2, it is not thought to arise as synchrotron radiation.
The form of the spectrum is very close to that 
given by Meisenheimer \& Heavens~(1987), who
used an analytic solution to the stationary diffusion/advection equation,
including synchrotron losses. Four free parameters are used to produce this 
fit: 
\begin{enumerate}
\item
the low frequency spectral index $\alpha=-0.25$, which corresponds to
taking $\tacc=\tesc/2$ 
\item
the characteristic synchrotron frequency emitted by an
electron 
of the maximum Lorentz factor 
as seen in the observers frame 
(taken to be $1.3\times10^{18}\,$Hz) 
\item
the spatial extent of the emitting region, which determines the 
position of the spectral break at roughly $5\times10^{12}\,$Hz 
\item
the absolute flux level.
\end{enumerate}
  
Since we restrict our model to the synchrotron emission of the accelerated
particles, it is not possible independently 
to constrain quantities such as the Doppler
boosting factor, or the magnetic field. 
Similarly, the frequency below which synchrotron self-absorption modifies the
optically thin spectrum is not constrained. 
Nevertheless, this model
of the synchrotron emission 
makes predictions concerning the spectral variability in each of the three
characteristic frequency ranges which can be identified in
Fig.~4. These ranges are generic features of any synchrotron model,
so that the predicted variability can easily 
be applied to the synchrotron emission of
other blazars. They are a) the low frequency region, where the particles have
not had time to cool before leaving the source (this is the region with
$\alpha=-0.25$ in Fig.~4, below the break at
$5\times10^{12}\,$Hz) b) the region between the break and the maximum flux,
where the particles have had time to cool, but where the cooling rate is always
much slower than the acceleration rate and the spectrum is close to 
$\alpha=-0.75$, and c) the region around and
above the flux maximum at roughly $10^{17}\,$Hz, where the acceleration rate is
comparable to the cooling rate.

\begin{figure}[t]
\vspace{0 cm}
\hspace{2 cm}
\epsfxsize=11. cm
\epsffile{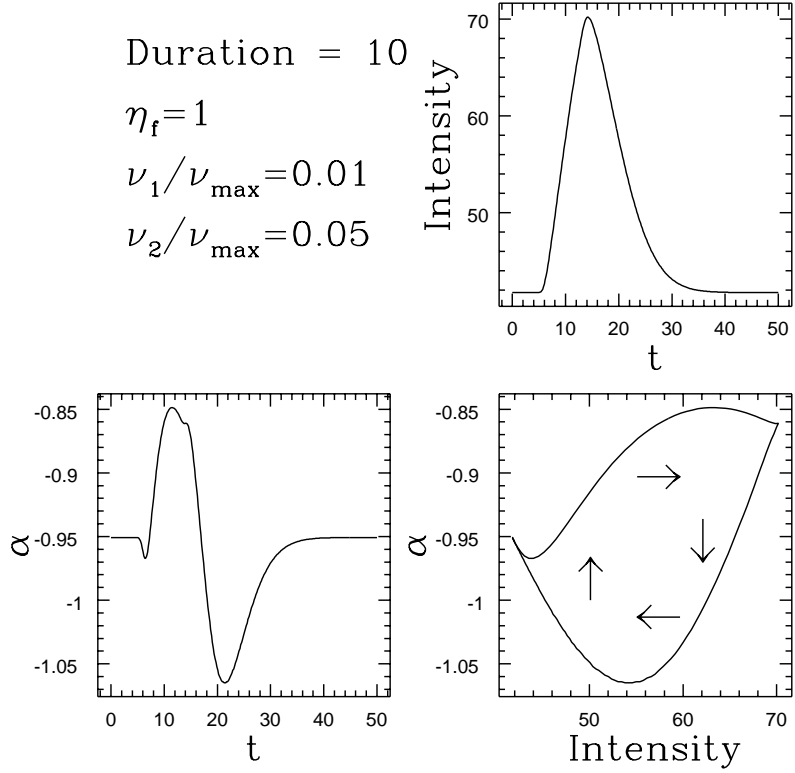}
\caption{The intensity and spectral index during the flare
described by Eq.~(\protect\ref{flareeq}),  
as a function of time at frequencies away from the high energy cut-off.
The loop in the $\alpha$ vs.\ intensity plot is followed in the clockwise
direction.}
\end{figure}

\begin{figure}[t]
\epsfxsize=11. cm
\hspace{2 cm}
\epsffile{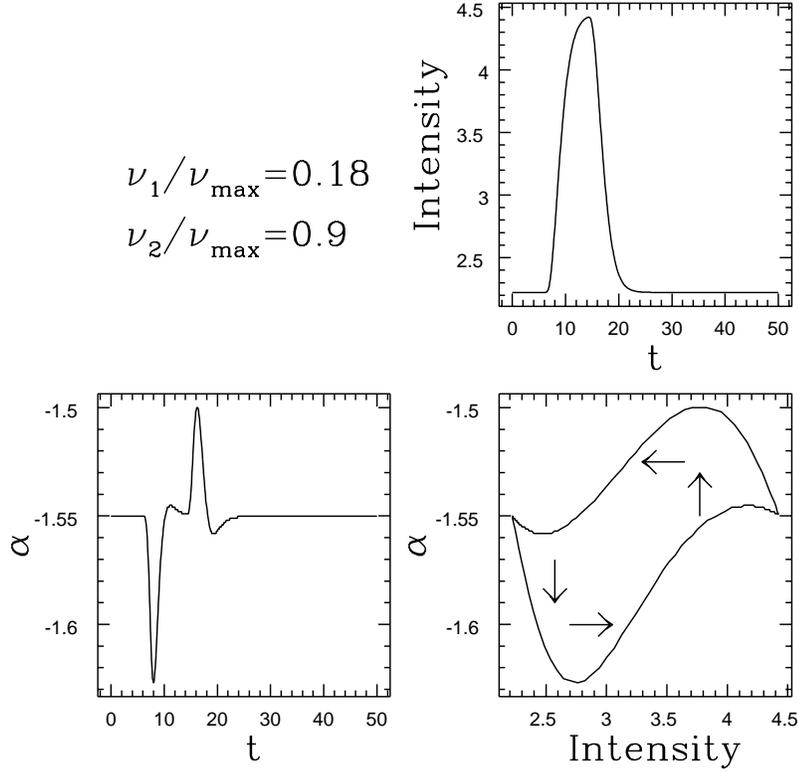}
\caption{\protect\label{loop2}
The intensity and spectral index during the flare
described by Eq.~(\protect\ref{flareeq}), as a function of time at 
low frequency. The loop in the $\protect\alpha$ vs.\ intensity plot
is followed in the anticlockwise direction.}
\end{figure}

Variability or flaring behaviour can arise for a number of reasons.
When the shock front overruns a region in the jet in which the local plasma
density is enhanced, the number of particles picked up and injected into the
acceleration process might be expected to increase. In addition, if the density
change is associated with a change in the magnetic field strength, the
acceleration timescale might also change, and, hence, the maximum frequency
of the emitted synchrotron radiation. Considering the case in which the
acceleration timescale remains constant, 
we can compute the emission in a straightforward manner.
An increase of the injection rate by a factor $1+\fflare$ for a time
$\tflare$ is found by setting
\eqb
Q(t)&=&Q_0\qquad {\rm for }\ t<0\ {\rm and}\ t>\tflare
\\
Q(t)&=&(1+\fflare) Q_0 \qquad {\rm for}\ 0<t<\tflare
\label{flareeq}
\eqe


Using $\fflare=1$, $\tflare=10\tacc$ and $u_{\rm s}=c/10$, we show
the resulting emission at a frequency $\nu=\nu_{\rm max}/100$ in Fig.~5.
In the case of Mkn~501, this corresponds to a frequency of about $10^{16}$ Hz,
which lies between the infra-red and X-ray regions where the spectral index
is close to $\alpha=-0.75$. Also shown in this figure is the temporal behaviour of the 
spectral index, as determined from the ratio of fluxes at $0.01\nu_{\rm {max}}$
and $0.05\nu_{\rm{max}}$, through the flare. When plotted against the flux at the
lower frequency, the spectral index exhibits a characteristic loop-like pattern, 
which is tracked in the clockwise sense by the system. This type of behaviour is well-known
and has been observed at different wavelengths in several sources e.g. OJ287 (Gear,
Robson \& Brown 1986), PKS 2155-304 (Sembay et al. 1993) 
and Mkn~421 (Takahashi et al. 1996).
It arises whenever the slope is controlled by synchrotron cooling so that information
about injection propagates from high to low energy (Tashiro et al. 1995).

If the system is observed closer to the maximum frequency, where the cooling and acceleration
times are equal, the picture changes. Here information about the occurrence of a flare
propagates from lower to higher energy, as particles are gradually accelerated
into the radiating window. Such behaviour is depicted in Fig.6, where the same flare
is shown at frequencies which are an order of magnitude higher than in Fig.5. In the case
of Mkn 501, the frequency range is close to $10^{18}$ Hz. This time the loop is traced anticlockwise. Such behaviour, although not as common, has occasionally
been observed in the case of PKS 2155-304 (Sembay et al. 1993).

\section{Summary}

In this paper we have presented a selective account of 
recent results on AGN variability within the context of a) the
homogeneous syncro-self-Compton and b) the diffusive particle acceleration.
We have shown that the SSC models give good overall fits to
the multiwavelength Mkn~421 and Mkn~501 spectra and can explain
the major flares of these objects such as the ones reported by Macomb et al
(1995) and Pian et al. (1998) respectively by increasing only
one parameter of the fit, namely the high energy cutoff of the 
injected electron distribution. This type of flare is especially prominent
at the high end of the photon distribution, i.e. in the X- and TeV regime, 
leaving other energy regimes (most notably the GeV gamma-rays)
practically unaffected, giving thus an explanation of 
why EGRET detected neither of these
two major outbursts. 

The very fast variation of the source Mkn~421, however, as reported by
Gaidos et al.\ (1996) poses a problem for the homogeneous SSC models:
in order for the models to satisfy 
simultaneously i) the high total luminosity,
ii) the very fast variability and iii) the transparency to TeV radiation 
(Bednarek \& Protheroe 1997b)
one needs either to invoke a high value of the Doppler boosting factor 
or to abandon the assumptions about a spherical source in favour of a laminar
source geometry. As shown above, this new manifestation
of the SSC model can provide us with good fits to the AGN observations both
in spectral and temporal behaviour (see, for examples, Figures 1-3).

This picture can lead naturally to the shock-in-jet model, i.e. to the picture of
a shock advancing down a jet, accelerating, at the same time, particles. Approaching
the acceleration by a first order Fermi scheme we have shown that one can get
once again remarkably good fits to the multiwavelength spectra of AGN at least from
the radio to the X-ray regime (since we have restricted our analysis only to the synchrotron
spectra--Fig.4). This approach improves upon the assumptions of homogeneous SSC
model, as presented in \S~2, 
mainly by replacing the instantaneous
electron injection with the concept of an acceleration timescale. It is therefore 
the interplay between the acceleration and energy loss timescales that provides
us with the 
different flare behaviours shown in Figures 5 and 6 (for a
more examples of this the reader is referred to Kirk, Rieger \& Mastichiadis   
1999).

\subsection*{Acknowledgments} 
AM would like to thank the organisers of the Workshop for their hospitality.
This work was supported by the European Commission under the TMR program,
contract number FMRX-CT98-0168. 

\section { References}

\reference
Ball L.T., Kirk J.G. 1992, ApJ 396, L39

\reference
Bednarek, W., Protheroe, R.J. 1997a, MNRAS 287, L9

\reference
Bednarek, W., Protheroe, R.J. 1997b, MNRAS 292, 646

\reference
Catanese, M. et al. 1997, ApJ 487, L143

\reference
Dermer, C.D., Schlickeiser, R., Mastichiadis, A. 1992, A\&A 256, L27

\reference
Dermer, C.D., Sturner, S.J., Schlickeiser, R. 1997, ApJS 109, 103

\reference
Gaidos, J.A. et al. 1996, Nature 383, 318

\reference
Gear, W.K., Robson, E.I., Brown, L.M.J. 1986, Nature 324, 546

\reference
Ghisellini, G., Maraschi, L., Dondi, L. 1996, A\&A Suppl 120C, 503

\reference
Hartman, R.C. et al. 1996, ApJ 461, 698

\reference
Inoue,S., Takahara, F. 1996, ApJ 463, 555

\reference
Jones, T.W., O'Dell, S.L., Stein, W.A. 1974, ApJ 188, 353

\reference
Kirk, J.G., Mastichiadis, A. 1997, in \lq Frontier Objects in Astrophysics and
Particle Physics\rq\ eds.: F. Giovannelli, G. Mannocchi, 
Conference Proceedings Vol.~57, page~263 
Italian Physical Society (Bologna)

\reference
Kirk, J.G., Melrose, D.B., Priest, E.R. Plasma astrophysics, eds.\ A.O. Benz, T.J.-L. Courvoisier, Springer, Berlin

\reference
Kirk, J.G., Rieger, F.M., Mastichiadis, A. 1998, A\&A 333, 452 (KRM)

\reference
Kirk, J.G., Rieger, F.M., Mastichiadis, A. 1999, 
to appear in proceedings of the conference \lq BL Lac Phenomenon\rq,
eds. L.O. Takalo, A. Sillanp\"a\"a, Turku 1998

\reference
Lin, Y.C. et al. 1992, ApJ 416, L53

\reference
Macomb, D.J. et al. 1995 ApJ 449, L99

\reference
Macomb, D.J. et al. 1996 ApJ 459, L111 (Erratum)

\reference
Mannheim, K. 1993, A\&A 269, 67

\reference
Maraschi L., Ghisellini G., Celotti A. 1992, ApJ 397, L5

\reference
Marscher, A.P., Gear, W.K. 1985, ApJ 298, 114

\reference
Marscher, A.P., Travis, J.P., 1996 A\&A Suppl 120C, 537

\reference
Mastichiadis A., Kirk, J.G., 1995 A\&A 295, 613 

\reference
Mastichiadis A., Kirk, J.G., 1997  A\&A 320, 19 

\reference
Meisenheimer, K., Heavens, A.F. 1987, MNRAS 225, 335

\reference
Pian, E. et al. 1998, ApJ 492, L17 

\reference
Punch, M. et al. 1992, Nature 358, 477

\reference
Quinn, J. et al. 1996, ApJ 456, 83

\reference
Salvati, M., Spada, M., Pacini, F. 1998, ApJ 495, L19

\reference 
Sembay, S. et al. 1993, ApJ 404, 112

\reference
Sikora, M., Begelman, M.C., Rees, M.J. 1994, ApJ 421, 153

\reference
Takahashi, T. et al. 1996, ApJ 470, L89

\reference
Tashiro, M. et al. 1995, PASJ 47, 131

\reference
Thompson, D.J. et al. 1995, ApJS 101, 259

\reference
Urry, C.M. \& Padovani, P. 1995, PASP 107, 803

\reference
Vermeulen, R.C. \& Cohen, M.H. 1994, ApJ 430, 467

\end{document}